\documentclass[showpacs,twocolumn]{revtex4}

\usepackage{color}

\usepackage{amsfonts}
\usepackage{amsmath}
\usepackage{amssymb}
\usepackage{graphicx}
\usepackage{subfigure}

\begin{document}

\title{Virial coefficients for Bose and Fermi trapped gases beyond the unitary limit: an S-Matrix approach}

\author{Edgar Marcelino}
\affiliation{
Institut f{\"u}r Theoretische Physik III, Ruhr-Universit{\"a}t Bochum, Universit{\"a}tsstra{\ss}e 150, DE-44801 Bochum, Germany}
\author{Andr\'e Nicolai}
\affiliation{
Centro Brasileiro de Pesquisas F\'{\i}sicas, Rua Dr. Xavier Sigaud 150, 22290-180 Rio de Janeiro, Rio de Janeiro, Brazil}
\author{Itzhak Roditi}
\affiliation{
Centro Brasileiro de Pesquisas F\'{\i}sicas, Rua Dr. Xavier Sigaud 150, 22290-180 Rio de Janeiro, Rio de Janeiro, Brazil}
\author{Andr\'e LeClair}
\affiliation{
Department of Physics, Cornell University, Ithaca, NY}

\begin{abstract}
\qquad
 
We study the virial expansion for three-dimensional Bose and Fermi gases at finite temperature using an approximation that only considers two-body processes and is valid for high temperatures and low densities. The first virial coefficients are computed and the second is exact.

The results are obtained for the full range of values of the scattering length and the unitary limit  is 
recovered  as a particular case.   A  weak coupling expansion is performed and the free case is also obtained as a proper limit.

The influence of an anisotropic harmonic trap is considered using the Local Density Approximation - LDA,  analytical results are obtained and the special case of the isotropic trap is discussed in
 detail.

\end{abstract}

\maketitle

\def\beq{\begin{equation}}
\def\eeq{\end{equation}}

\section{Introduction}

\qquad

The advances in experimental results and simulations on cold atoms \cite{Cold_atoms1,Cold_atoms2,Cold_atoms3,Cold_atoms4,Cold_atoms5,Cold_atoms6,MC1,MC2,MC3,MC4,Review1,Review2} requires new methods for theorists to study these systems and explore similar ones. Analytical methods continue 
 to be a powerful tool to explore these systems, although they generally provide approximate results, in comparison with numerical methods  \cite{Harmonic}.

This work uses a formalism for  Statistical Mechanics based on the S-Matrix \cite{Dashen}. It provides an expression of the free energy at finite temperature and density built on  an integral equation of the pseudo-energy with a kernel based on the logarithm of the 2-body S-matrix at zero temperature.
This integral equation is quite similar to the Yang-Yang  equations used in the Thermodynamical Bethe Ansatz - TBA \cite{YangYang}. 

The method is a ``foam diagram''  approximation which is valid for high temperatures and low densities and considers only contributions from two-body processes to the free energy.  It is explained in \cite{PyeTon} and has been already used to study the thermodynamical and critical properties of quantum gases in two and three dimensions in the unitary limit \cite{PyeTonUnitary1,PyeTonUnitary2} and beyond the unitary limit in three dimensions \cite{Out_unitarity}.  In \cite{Leclair_viscosity} the method was used to calculate the ratio of the viscosity to entropy density and the results were in well agreement with experimental data \cite{Exp_viscosity}.

In \cite{Out_unitarity} it was shown how  this method may be used to obtain the coefficients of the virial expansion for quantum gases and the first four virial coefficients were calculated in three dimensions in the unitary limit. The second coefficient provided by this method is exact and agrees with the result  in \cite{HoMu}.  The third one in the unitary limit does not agree with the exact value obtained in \cite{b3Theory,Leyronas} where three-body processes were considered since the three-body processes are neglected in our approximation. 
What is new about the present work is that we extend this analysis to arbitrary scattering 
length;   the unitary limit is obtained as the scattering length goes to infinity.   
These remarks apply to both bosons and fermions,  and both cases are considered here,
whereas the literature mainly deals with fermions in the unitary limit.   

Since Feshbach Resonance experiments allow to adjust the scattering length to any finite value there is no reason to study  only the unitary limit, in which the scattering length diverges. 
Here we calculate the first three virial coefficients for both Bose and Fermi gases in three dimensions for different values of the dimensionless ratio  
$$\alpha=\frac{\lambda_{T}}{a}, $$where $\lambda_{T}=\sqrt{\frac{2 \pi}{mT}}$ is the De Broglie thermal wave length and $a$ is the scattering length. The unitary limit results obtained in  \cite{Out_unitarity} are recovered in the proper limit and also  the free case where the scattering length is tuned to zero. For large positive scattering length molecules are formed and this is not considered in this work.  However  in the  ``upper branch'' \cite{Up1,Up2,Up3,Up4,Up5,Up6,Up7} there are no molecules and our formalism may be applied. This situation is studied here because we consider the possibility that a Bose gas may stay in a metastable state before undergoing  mechanical collapse \cite{Ketterle}.   Virial coefficients on the upper branch have not been considered before.   

Analytical expressions are obtained for a  weak coupling expansion and they are compared to the previous results.  The second virial coefficient is the only exact one (besides the first one) for the same reasons as in  in \cite{Out_unitarity} and this will be discussed here.  Finally the influence of a harmonic trap on the virial coefficients will be studied using the Local Density Approximation (LDA) and analytic results will be obtained for the case of an anisotropic harmonic trap.  The particular case of the isotropic trap will be discussed and some plots will be showed. 

In the next section we present a brief summary  of the formalism (for more details see \cite{PyeTon}), 
the  actions of our physical systems,  and the conventions used in this paper. In section III we derive the expression of the virial coefficients in terms of the two-body kernel of the theory in the foam diagram approximation in a different way than in  \cite{Out_unitarity}. In section IV we obtain the first four virial coefficients of a Bose and a Fermi gas in three dimensions in terms of the ratio $\alpha=\frac{\lambda_{T}}{a}$ and discuss these results. In section V we perform a weak coupling expansion and obtain analytical expressions for the virial coefficients in this situation, then we compare these results with the previous ones obtained in section IV. In section VI we study the influence of a trap on the virial coefficients using the Local Density Approximation.

\section{Formalism and conventions}

\def\om#1{\omega_#1}
\def\kvec{{\bf k}}
\def\omk{\omega_{\kvec}}
\def\inv#1{\frac{1}{#1}}
\def\d{\partial}

\qquad

In this section we review the main result in \cite{PyeTon}.
The formalism is developed starting with the  fundamental formula \cite{Dashen}
for the partition function $Z$:  
\beq
\label{Dashen}  
Z = Z_0 + \frac{1}{2\pi}  \int dE\, e^{-\beta E} \, {\rm Tr}  \, {\rm Im}  \d_E  \log \hat{S} (E) 
\eeq
where  
$Z_0$ is the partition function for the free theory,  $\beta = 1/T$  is the inverse temperature,  and 
$\hat{S}$  is the S-matrix operator
in scattering theory.   
A considerable amount of work is needed to turn this into a useful expression,
such as the cluster expansion for the $S$-matrix.    The result is a diagrammatic 
expansion for the free energy,  not to be confused with finite temperature 
Feynman diagrams which are perturbative in the coupling.   Rather,   
diagrams consist of vertices with $2n$ legs  which correspond to the logarithm of
the $n$-particle to $n$ particle $S$-matrix at zero temperature,  up to some
kinematical factors.     (For non-relativistic theories
there is no particle production.)    These vertices are connected by lines 
which are the occupation numbers 
\begin{equation}
\label{fo} 
f_{0}(\mathbf{k})= \frac {z} {e^{\beta \omk}-s\, z},
\end{equation}
where $s$ is a statistical parameter (1 for bosons and -1 for fermions) and  $z=e^{\beta \mu}$ is the fugacity, where $\mu$ is the chemical potential.   Here we are only considering 
non-relativistic theories where $\omk = \kvec^2/2m$,  $m$ being the mass of the particles.   

\begin{figure}[tbp]
\centering    
\includegraphics[width=0.5\textwidth]{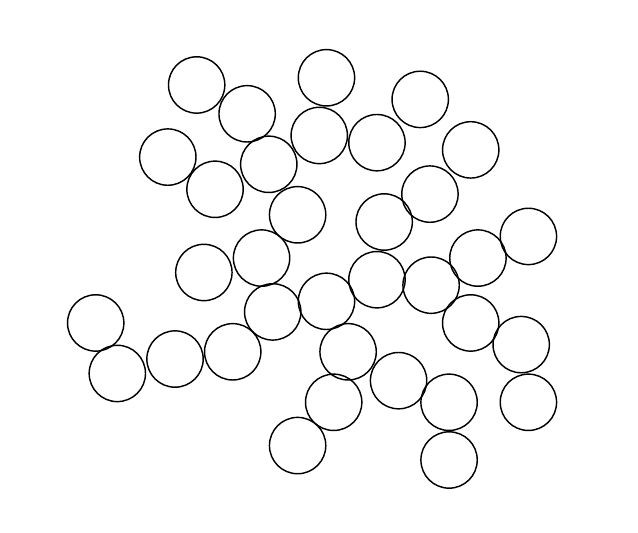}
\caption{Foam diagrams}
\label{foam}
\end{figure}

For the non-relativistic theories we will consider,  the two body $S$-matrix,   i.e. 
$n=2$,  can be calculated exactly,  i.e. to all orders in the coupling,    thus this formalism 
captures   some non-perturbative aspects.  However the vertices for $n>2$ are difficult to calculate.
Thus we consider the approximation where we consider only diagrams involving $2$-body
scattering.    These are the diagrams shown in Figure \ref{foam},   i.e. the ``foam"  diagrams.   
This infinite class of diagrams can be summed up leading to an integral equation we describe below.  

In this formalism,   the filling fractions,  or occupation numbers,  are parametrized in 
terms of a pseudo-energy $\varepsilon (\kvec)$ which has the same form as the free theory,  i.e. 
the density has the following expression: 
\begin{equation}   
 \label{number}
n=\int \frac{d^3 \mathbf{k}} {(2 \pi)^3} \frac{1} {e^{\beta \varepsilon(\mathbf{k})}-s},
\end{equation}
The summation of all foam diagrams leads to an integral equation satisfied by
$\varepsilon$ which we now describe.  
It is convenient to define 
\begin{equation} 
  \label{y}
y(\mathbf{k})=e^{-\beta  \varepsilon(\mathbf{k}  - \omk + \mu)}
\end{equation}
It satisfies the integral equation 
\begin{equation} \label{integral equation}
y(\mathbf{k})=1+ \beta  \int   \frac{d^3 \kvec'}{(2\pi )^3}  \, G(\kvec - \kvec' ) 
\frac{z}{e^{\beta \omega_{\kvec'}} - s\, z \, y(\kvec') }
\end{equation}
The kernel $G$ is the logarithm of the 2-body $S$-matrix multiplied by 
a kinematical factor which will be specified below.   
The pseudo-energy $\varepsilon$  may be interpreted as a self-energy 
correction in the presence of all the particles of the gas that takes into account
multiple scatterings,   however our formalism is different than others in
the literature that can also be interpreted as self-energies.   
It is different than the ``self-consistent T-matrix approximation" for instance,
since the latter does not involve our kernel $G$.     
The free energy is given by the following formula: 
\begin{equation} 
  \label{free energy}
F=-\frac {1} {\beta} \int \frac {d^3 \mathbf{k}} {(2 \pi)^3} \left[ -s\, \log(1-se^{-\beta \varepsilon (\kvec)}) - \frac {1} {2} \left( \frac{1-y(\kvec) ^{-1}} {e^{\beta \varepsilon(\kvec)}-s} \right) \right].
\end{equation}

We will study both Bose and Fermi gases.    The Bose gas will be described by the following action:
\begin{equation}
S=\int d^{3} \mathbf{x} dt \left(i \phi^{\dagger} \partial_{t} \phi-\frac{{\nabla \phi}^{2}}{2m}-\frac{g}{2} \left(\phi^{\dagger} \phi \right)^{2} \right)
\end{equation}
and the fermion gas by:
\begin{equation}
S=\int d^{3} \mathbf{x} dt \left(\sum_{\alpha=\uparrow,\downarrow}i \psi_{\alpha}^{\dagger}\partial_{t} \psi_{\alpha}-\frac{{\nabla \psi_{\alpha}}^{2}}{2m}-g \psi_{\uparrow}^{\dagger}\psi_{\uparrow}\psi_{\downarrow}^{\dagger}\psi_{\downarrow} \right).
\end{equation}
 The  renormalized coupling constant is  given by:
\begin{equation}
\frac{1}{g_{R}}=\frac{1}{g}+\frac{m \Lambda}{2 \pi^{2}},
\end{equation}
where $\Lambda$ is the momentum cutoff introduced to regularize loop integrals.  The scattering length is  related to the renormalized coupling constant by:
\begin{equation}
a=\frac{mg_{R}}{4 \pi}.
\end{equation}
In the unitary limit,  the scattering length $a \to \pm \infty$ and the theory is scale invariant,
i.e. at the renormalization group fixed point.   

The 2-body $S$ matrix  is 
\beq
\label{Smatrix}
S_{\rm matrix}  ( |\kvec - \kvec' |) =  \frac{ 8 \pi/mg_R - i |\kvec - \kvec'|}{8 \pi/mg_R + i |\kvec - \kvec'|}
\eeq
and in the unitary limit  it simply  equals $-1$.  
The kernel that follows from this $S$-matrix is: 
\begin{equation}
 \label{kernel}
G(\mathbf{k},\mathbf{k'})=-\frac {16 \pi \sigma} {m. |\mathbf{k}-\mathbf{k'}|} \arctan  \left( \frac {mg_{R}.| \mathbf{k}-\mathbf{k'} |} {8 \pi} \right),
\end{equation}
where the factor $\sigma$ that appears in (\ref{kernel}) is $\sigma= 1/2$ 
 for fermions and $\sigma=1$ for bosons. 
 Note that in the unitary limit the kernel remains a non-constant but much simpler function, 
 namely as $a\to \mp \infty$ one has
 \beq
 \label{Gunitary}
 G ( \kvec  ) = \pm \frac{8 \pi^2 \sigma}{m |\kvec|} .
 \eeq
  
   In the
present work,  the first virial coefficients will be obtained for any value of the scattering length and the unitary limit will be recovered as $\alpha \to 0$.

\section{Expressions for the virial coefficients in the foam diagram approximation}

\qquad

The virial coefficients $b_{i}$ may be defined by the following expression:
\begin{equation} \label{free_energy_virial}
F=-\frac{1}{\beta \lambda_{T}^{3}} \sum_{n=1}^{\infty} b_{n}z^{n}.
\end{equation}
It is convenient to define a  dimensionless scaling function $q$ for  the density of particles
which is only a function of $\mu/T$ and $\alpha = \lambda_T /a$,  as follows:
\begin{equation} \label{n scale}
q=n\lambda_{T}^{3}.
\end{equation}
Recalling  that $n=-\frac{\partial F}{\partial \mu}$ one obtains:
\begin{equation}  \label{virial}
q=  \sum_{n=1}^{\infty} n\, b_{n}z^{n}.
\end{equation}

Substituting (\ref{number}) in (\ref{n scale}) and using (\ref{y}), it is possible to expand $q$ as
follows 
\begin{multline}    \label{qexpansion}
q=\left( \frac {1} {2 \pi mT} \right)^{3/2} \int d^{3} \mathbf{k} \,  z\,y(\mathbf{k}) e^{-\beta \omk } [1+ szy(\mathbf{k}) e^{- \beta \omk }+ \\
+z^2 y^2 (\mathbf{k}) e^{-2 \beta\omk }+...]  
\end{multline}

Using (\ref{y}) and (\ref{integral equation}) it is possible to expand $y(\mathbf{k})$, 

\begin{multline}   
  \label{yexpansion}
y(\mathbf{k})=1+\frac {\beta} {(2 \pi)^3} \int d^3 \mathbf{k}' ~ G({\mathbf{k},\mathbf{k}'})\, z\,
e^{-\beta \omega_{\kvec'}}  [ 1+  \\
+s\,z\,y(\mathbf{k}')\, e^{- \beta \omega_{\kvec'}}
+z^2 y^2 e^{-\beta \omega_{\kvec'}} +... ].  
\end{multline}
Now, using (\ref{qexpansion}) and (\ref{yexpansion}) we can express  the scaling  function $q$ in terms of the fugacity.  Comparing to (\ref{virial})  one can then obtain expressions for  the virial coefficients.  The first three are:

\begin{equation}
(2 \pi mT)^{3/2} \,  b_{1}= \int  {d^{3} \mathbf{k}} \,  e^{-\beta \omk},
\end{equation}

\begin{multline}
2(2 \pi mT)^{3/2} \,  b_{2}=s \int  {d^{3} \mathbf{k}} \, e^{-2 \beta \omk} +  \\
+\frac {\beta} {{(2 \pi)}^3} \int  d^{3}\,  \mathbf{k}  d^{3} \mathbf{k}' \, e^{-\beta \omk} e^{-\beta \omega_{\kvec'} } \, G(\mathbf{k},\mathbf{k}'),
\end{multline}

\begin{multline}
3(2 \pi mT)^{3/2} \,  b_{3}=\int  {d^{3} \mathbf{k}} \, e^{- 3 \beta \omk} + \\
+s\frac {2 \beta} {{(2 \pi)}^3} 
\int  d^{3} \mathbf{k}  \, d^{3} \mathbf{k}' \, e^{-2\beta \omk } e^{-\beta  \omega_{\kvec'} } \, G (\mathbf{k},\mathbf{k}')+ \\
+ \frac {\beta s} {{(2 \pi)}^3} \int  d^{3} \mathbf{k}  \, d^{3} \mathbf{k}'\,  e^{-\beta \omk } e^{-2 \beta \omega_{\kvec'} } \,   
G (\mathbf{k},\mathbf{k}'),
\end{multline}

Performing the integrals that do not depend on the kernel we obtain the simpler expressions: 
 \begin{equation} \label{b1}
 b_1=1,
 \end{equation}
 
\begin{multline}     \label{b2}
2b_{2}=\frac{s} {2 ^{3/2}}+\frac {\beta} {{(2 \pi)}^3 (2 \pi mT)^{3/2}}  \\
\int  d^{3} \mathbf{k} \,  d^{3} \mathbf{k}' \, e^{-\beta\omk } e^{-\beta \omega_{\kvec'} } \, G(\mathbf{k},\mathbf{k}'),
\end{multline}

\begin{multline} \label{b3}
3b_{3} =\frac{1} {3^{3/2}}+\frac{3 \beta s} {(2 \pi mT)^{3/2} (2 \pi)^3} \\
\int  d^{3} \mathbf{k}  \, d^{3} \mathbf{k}' \, e^{-2 \beta \omk} \,  e^{-\beta \omega_{\kvec'}} \,  G(\mathbf{k},\mathbf{k}').
\end{multline}

The above derivation of these expressions  is slightly different than the one in
 \cite{Out_unitarity},  
 since it is  not necessary to consider each diagram and find its contributions to the virial coefficients.  
The method presented in this paper automatically considers the foam diagram approximation because of the use of the integral equation (\ref{integral equation});   the fact that the results here are in agreement with \cite{Out_unitarity} shows the consistency  of the formalism presented in \cite{PyeTon}.

The second virial coefficient is exact,  as was shown previously in the unitary limit \cite{Out_unitarity}.  
Higher coefficients have contributions from two body scattering,  but since we
do not take into account primitive higher body processes they give less precise results. 
For completeness,  we give the contribution to $b_4$ from 2-body interactions: 
\begin{multline}    \label{b4}
4b_{4}=\frac{s} {4^{3/2}} +  \\
 +\frac {4 \beta} {(2 \pi mT)^{3/2} (2 \pi)^{3}} 
\int e^{-3 \beta \omk }  e^{- \beta \omega_{\kvec'} } G(\mathbf{k},\mathbf{k}')\,  d^{3}\mathbf{k} \, d^{3}\mathbf{k'}+ \\
+\frac {2 \beta} {(2 \pi)^{3} (2 \pi mT)^{3/2}}  \int e^{-2 \beta \omk}  e^{-2 \beta \omega_{\kvec'}} 
G(\mathbf{k},\mathbf{k}')\,  d^{3}\mathbf{k}\,  d^{3}\mathbf{k'}+       \\
+ \frac {{2 \beta}^2 s} {{(2 \pi mT)}^{3/2} {(2 \pi)}^{6}}  \int e^{- \beta \omk} 
 e^{-2 \beta \omega_{\kvec'} }  e^{- \beta  \omega_{\kvec''} } G(\mathbf{k},\mathbf{k}')  \\
 G(\mathbf{k}',\mathbf{k}'') \, d^{3}\mathbf{k} \, d^{3}\mathbf{k'}
 d^{3}\mathbf{k''}.
 \end{multline}

\section{The results for the first virial coefficients}

\qquad

Substituting the kernel (\ref{kernel}) in equations (\ref{b1}), (\ref{b2}), (\ref{b3}) and performing  the angular parts of the integrals,  it is possible to write the first virial coefficients in terms of the
ratio  $\alpha= \lambda_{T}/a$.  The results are the following:
\begin{equation}
b_{1}=1
\end{equation}
\begin{equation} \label{b2s}
b_{2}=\frac{s \sqrt{2}}{8}-\frac{2 \sqrt{2} \sigma}{\pi^{2}} \int_{0}^{\infty} v\, e^{-\frac{v^{2}}{2 \pi}}
\arctan \left( \frac{v}{\alpha} \right) dv,
\end{equation}
\begin{equation} \label{b3s}
b_{3}=\frac{ \sqrt{3}}{27}-\frac{16 \sqrt{3} \sigma s}{9 \pi^{2}} \int_{0}^{\infty} v\, e^{-\frac{2v^{2}}{3 \pi}}\arctan \left( \frac{v}{\alpha} \right) dv.
\end{equation}

For the sake of completeness we also give the expression of the fourth coefficient,
\begin{align}
b_{4}&=\frac{s}{32}-\frac{2 \sigma}{\pi^{2}} \int_{0}^{\infty}ve^{-\frac{3v^{2}}{4 \pi}} \arctan \left( \frac{v}{\alpha} \right) dv\\\notag &-\frac{\sigma}{\pi^{2}} \int_{0}^{\infty}ve^{-\frac{v^{2}}{\pi}} \arctan \left( \frac{v}{\alpha} \right) dv \\\notag
&+\frac{128 \sigma^{2} s}{\pi^{5/2}} \int_{0}^{\infty} \int_{0}^{\infty} \int_{0}^{\infty} e^{-(u^{2}+v^{2}+4w^{2})} \\\notag &\times\sinh(uw) \sinh(vw) \arctan \left( \frac{u \sqrt{\pi}}{\alpha} \right)\\\notag &\times \arctan \left( \frac{v \sqrt{\pi}}{\alpha} \right)\, uv\, du\, dv\, dw.
\end{align}

In the unitary limit,  the above integrals can be performed analytically \cite{Out_unitarity}.  
For a finite scattering length $a$,  the integrals can only be done numerically.

\begin{figure}[tbp]
\centering    
\includegraphics[width=0.5\textwidth]{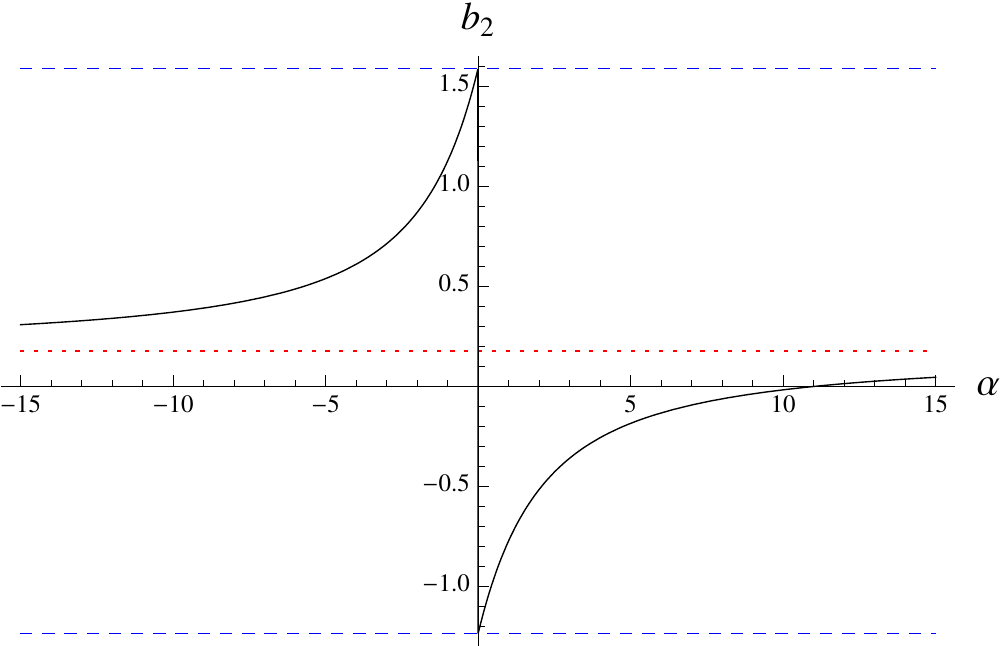}
\caption{(Color Online): Second virial coefficient against the ratio between the thermal wave length and the scattering length: $b_{2}$ $\times$ $\alpha=\frac{\lambda_{T}}{a}$ for bosons (black). The values of the unitary limit ($\alpha \rightarrow 0^{\pm}$) obtained in \cite{Out_unitarity} are represented by the dashed (blue) lines and the value of the free case 5($g =0 \Rightarrow \alpha \rightarrow \pm \infty$) is represented by the dotted (red) line.}
\label{b2BxT}
\end{figure}

\begin{figure}[tbp]
\centering    
\includegraphics[width=0.5\textwidth]{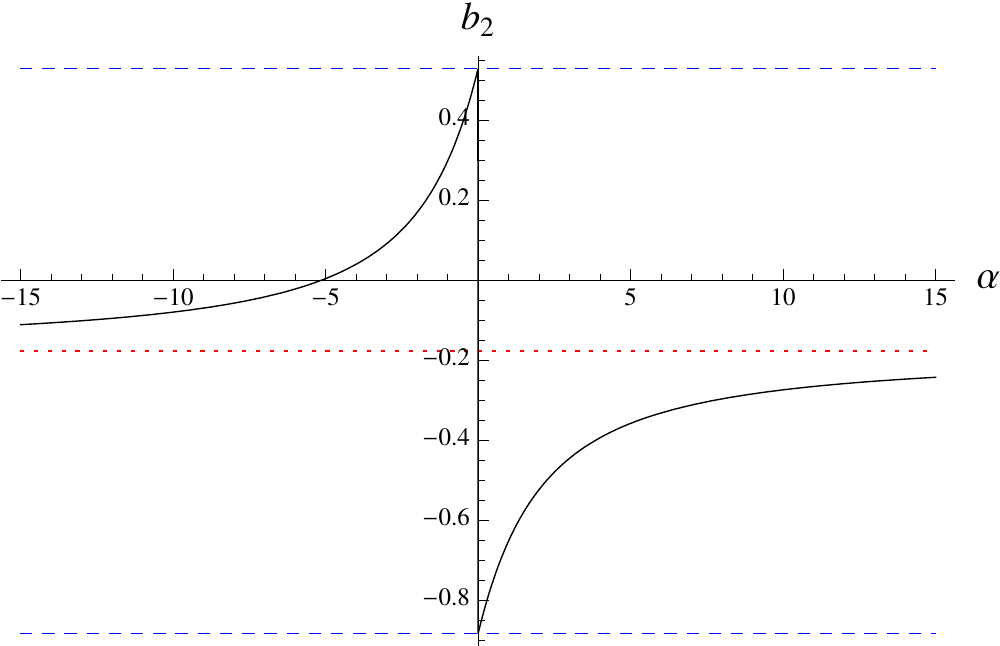}
\caption{(Color Online): Second virial coefficient against the ratio between the thermal wave length and the scattering length: $b_{2}$ $\times$ $\alpha=\frac{\lambda_{T}}{a}$ for fermions (black). The values of the unitary limit ($\alpha \rightarrow 0^{\pm}$) obtained in \cite{Out_unitarity} are represented by the dashed (blue) lines and the value of the free case ($g =0 \Rightarrow \alpha \rightarrow \pm \infty$) is represented by the dotted (red) line.}
\label{b2FxT}
\end{figure}

\begin{figure}[tbp]
\centering    
\includegraphics[width=0.5\textwidth]{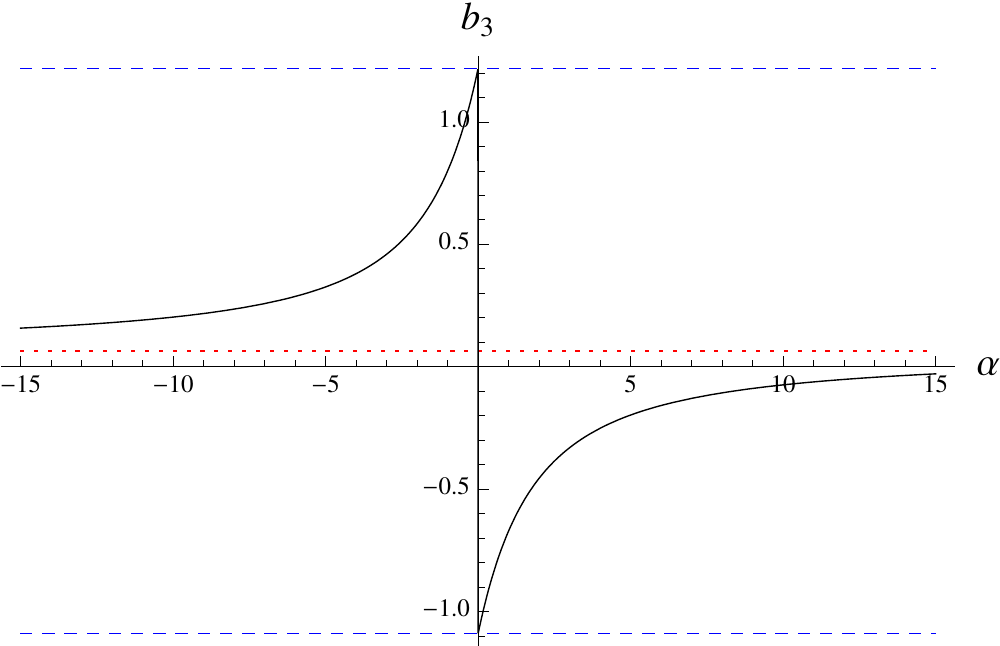}
\caption{(Color Online): Third virial coefficient against the ratio between the thermal wave length and the scattering length: $b_{3}$ $\times$ $\alpha=\frac{\lambda_{T}}{a}$ for bosons (black). The values of the unitary limit ($\alpha \rightarrow 0^{\pm}$) obtained in \cite{Out_unitarity} are represented by the dashed (blue) lines and the value of the free case ($g =0 \Rightarrow \alpha \rightarrow \pm \infty$) is represented by the dotted (red) line.}
\label{b3BxT}
\end{figure}

\begin{figure}[tbp]
\centering    
\includegraphics[width=0.5\textwidth]{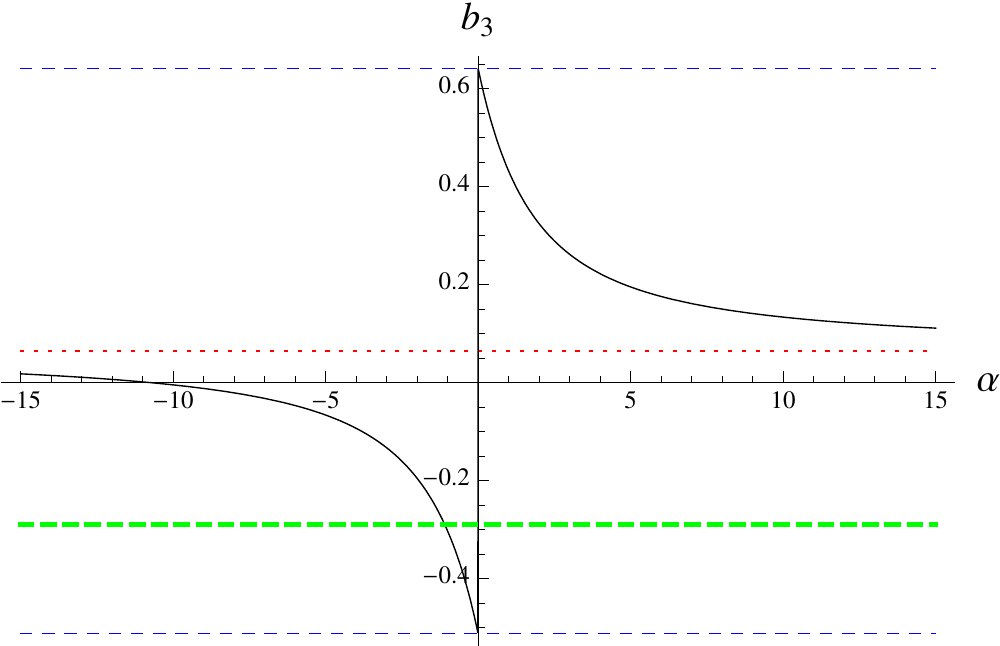}
\caption{(Color Online): Third virial coefficient against the ratio between the thermal wave length and the scattering length: $b_{3}$ $\times$ $\alpha=\frac{\lambda_{T}}{a}$ for fermions (black). The values of the unitary limit ($\alpha \rightarrow 0^{\pm}$) obtained in \cite{Out_unitarity} are represented by the dashed (blue) lines and the value of the free case ($g =0 \Rightarrow \alpha \rightarrow \pm \infty$) is represented by the dotted (red) line. The exact result for the untarity limit with infinite negative scattering length obtained in \cite{Leyronas} is represented by a thick dashed (green) line.}
\label{b3FxT}
\end{figure}

The second virial coefficient as a function of  $\alpha$ is plotted in Figure \ref{b2BxT} for bosons and in Figure \ref{b2FxT} for fermions, Figures \ref{b3BxT} and \ref{b3FxT} show the third virial coefficient for bosons and fermions respectively. The values of these coefficients in  the free case and in the unitary limit are also indicated in these  figures  with dotted and dashed lines respectively 
and one sees they are recovered in the proper limits $a\to 0$ and $a\to \infty$.    Note that 
both $b_2$ and $b_3$ flip sign as one passes through the unitary limit and the scattering 
length changes from $+ \infty$  to $- \infty$.    This is due to the exclusion of the bound
state for both $b_2$ and $b_3$  since  our $b_3$ is still only based on 2-body physics.

The figures \ref{b2BxT}, \ref{b2FxT}, \ref{b3BxT} and \ref{b3FxT} show that the second and third virial coefficients are bounded by the values of the unitary limit case (when $\alpha \rightarrow 0$, the dashed lines) in the foam diagram approximation.  When $g \rightarrow 0^{\pm} \Rightarrow \alpha \rightarrow \pm \infty $ (the dotted lines),  the free case is always recovered as expected. The exact results for the values of the second coefficient  in the unitary limit  are  also properly recovered for Bose and Fermi gases \cite{HoMu,b3Theory,Leyronas,Kaplan,Out_unitarity}.  The expression (\ref{b2s}) for fermions ($s=-1$, $\sigma=\frac{1}{2}$) is the same as the exact one obtained in \cite{Leyronas}, 
up to an integration by parts,   thus our results  for the second virial coefficient are exact for the whole range  of $\alpha$, as expected from our formalism.   

The results in the unitary limit for the third coefficient obtained in \cite{Out_unitarity} are recovered as expected.   As discussed previously,   they differ from the exact ones from \cite{Leyronas,Kaplan,Castin} since the three-body processes are not considered in our approximation. As the ratio $\alpha$ increases, the interaction effects decrease and our results should become closer to the correct ones.  
  The Figure \ref{b3FxT}  indeed shows  that for $\alpha$ sufficiently large the results obtained in \cite{Leyronas} are nearly recovered, the exact value for this coefficient obtained in \cite{Leyronas} for the unitary limit with large negative scattering length is also showed in this figure with a thick dashed line.  
  
\section{Weak coupling expansion}

\qquad

Expressions  (\ref{b2s}) and (\ref{b3s}) give the virial coefficients in terms of the ratio $\alpha=\frac{\lambda_{T}}{a}$.  In  this section we  perform an  expansion for large values of $\alpha$, which means that $\sqrt{T}g_{R}<<1$. Since the temperature can not be too small because we are under the foam diagram approximation,  the coupling constant $g_{R}$ should be very small in order for  this expansion be valid. 

One can simply expand the arc-tangent function in expressions  (\ref{b2s}) and (\ref{b3s}) in a Taylor series, truncate it to  the first degree term of $\frac{1}{\alpha}$,  and perform  the integrals analytically.  
Performing this expansion, one obtains 
\begin{equation} \label{wb2s}
b_{2}=\frac{s \sqrt{2}}{8}-\frac{2 \sigma}{\alpha} 
\end{equation}
and
\begin{equation} \label{wb3s}
b_{3}=\frac{ \sqrt{3}}{27}-\frac{ \sqrt{2} \sigma s}{\alpha}.
\end{equation}

\begin{figure}[tbp]
\centering    
\includegraphics[width=0.5\textwidth]{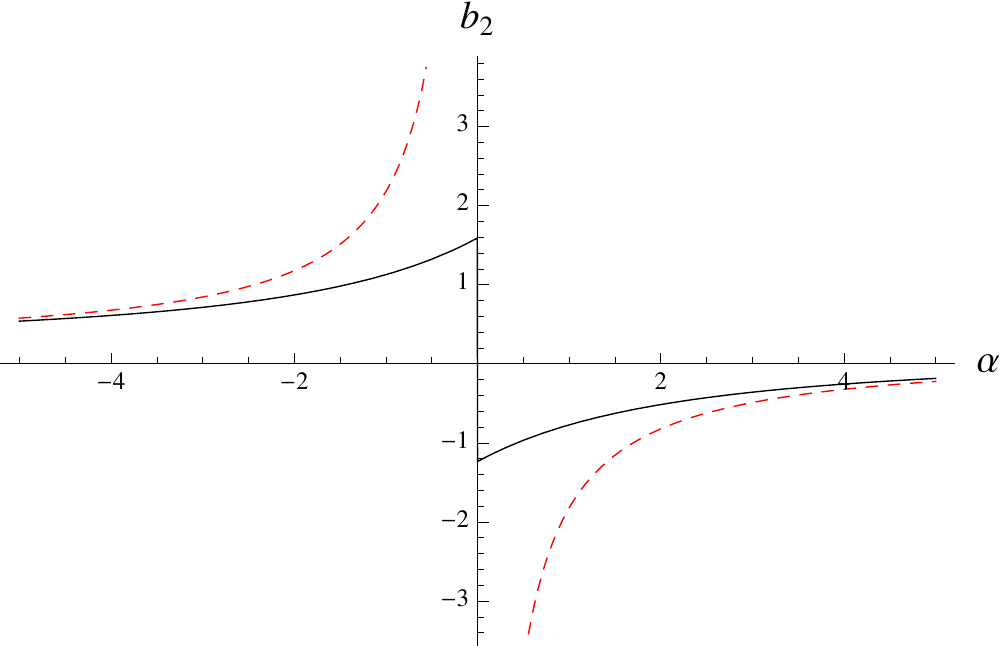}
\caption{(Color Online): Second virial coefficient against the ratio between the thermal wave length and the scattering length: $b_{2}$ $\times$ $\alpha=\frac{\lambda_{T}}{a}$ for bosons (black). The dashed (red) line shows the same result obtained with the expression of the weak coupling expansion.}
\label{wbb2}
\end{figure}

\begin{figure}[tbp]
\centering    
\includegraphics[width=0.5\textwidth]{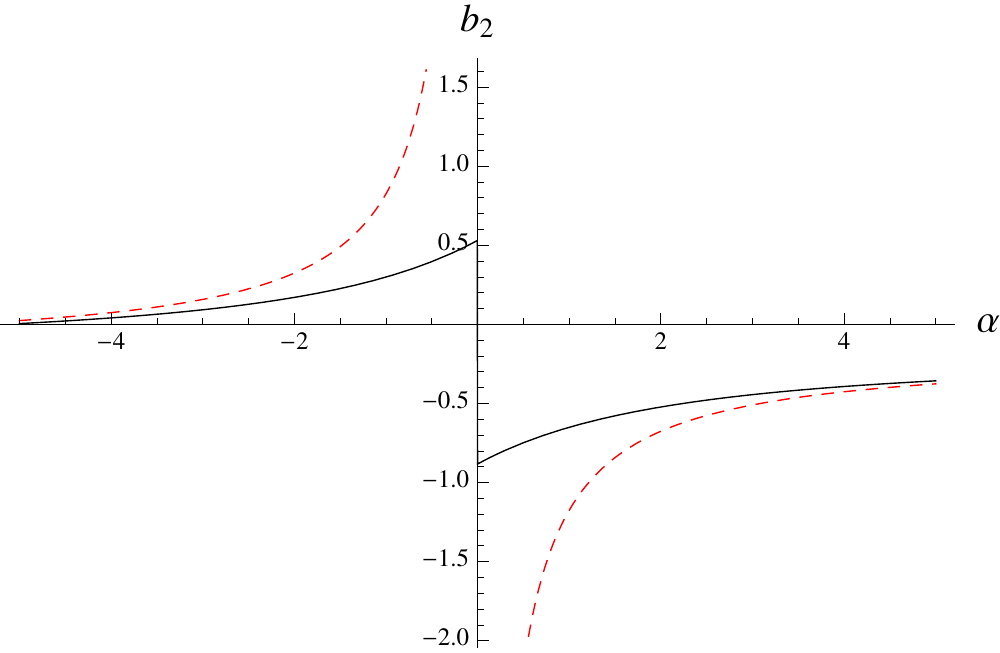}
\caption{(Color Online): Second virial coefficient against the ratio between the thermal wave length and the scattering length: $b_{2}$ $\times$ $\alpha=\frac{\lambda_{T}}{a}$ for fermions (black). The dashed (red) line shows the same result obtained with the expression of the weak coupling expansion.}
\label{wfb2}
\end{figure}

\begin{figure}[tbp]
\centering    
\includegraphics[width=0.5\textwidth]{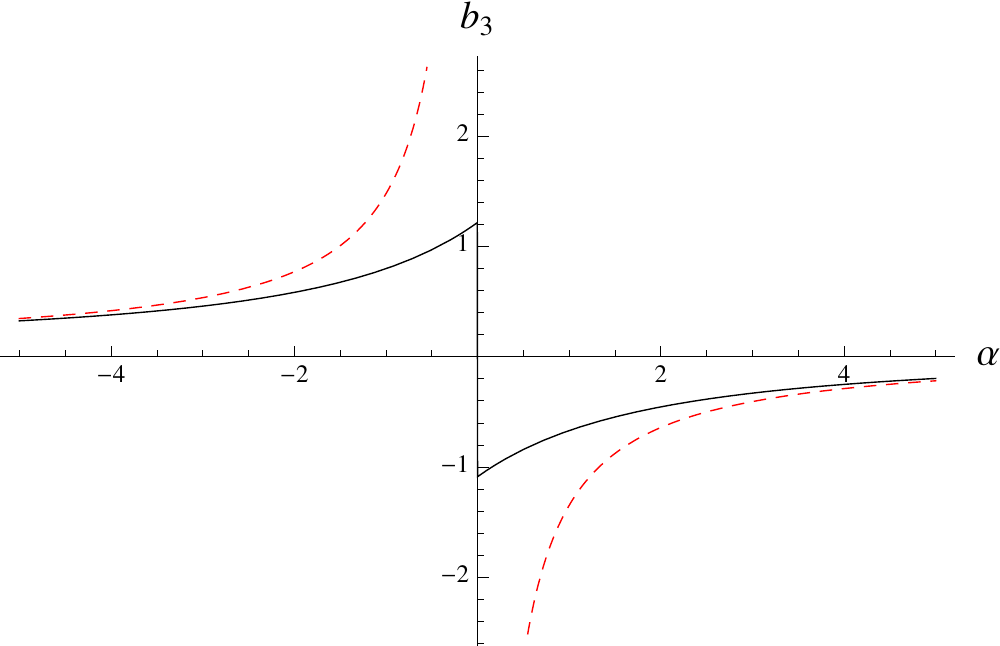}
\caption{(Color Online): Third virial coefficient against the ratio between the thermal wave length and the scattering length: $b_{2}$ $\times$ $\alpha=\frac{\lambda_{T}}{a}$ for bosons (black). The dashed (red) line shows the same result obtained with the expression of the weak coupling expansion.}
\label{wbb3}
\end{figure}

\begin{figure}[tbp]
\centering    
\includegraphics[width=0.5\textwidth]{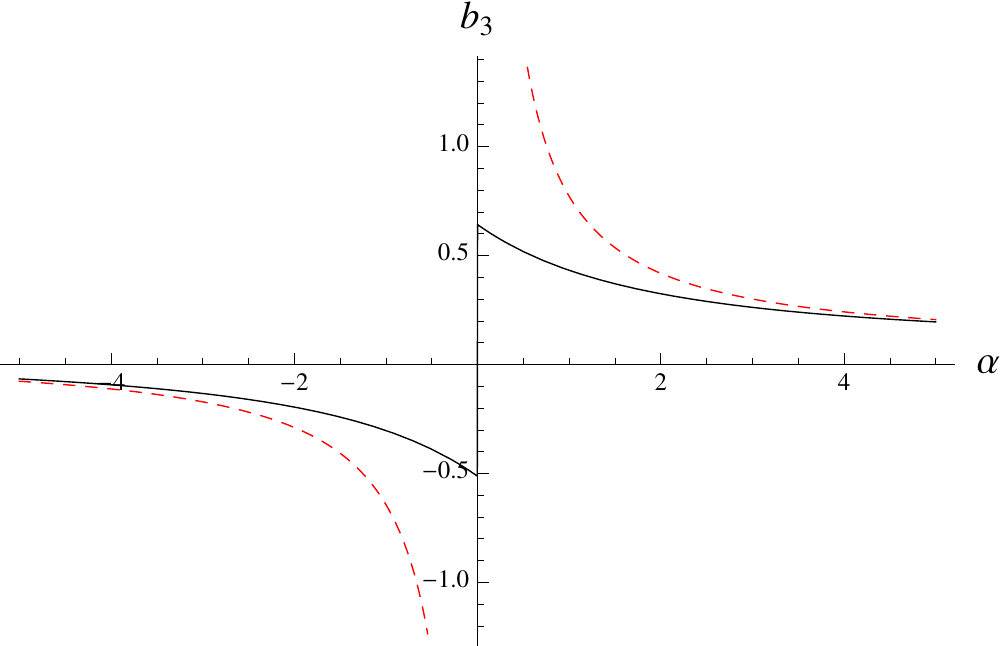}
\caption{(Color Online): Third virial coefficient against the ratio between the thermal wave length and the scattering length: $b_{2}$ $\times$ $\alpha=\frac{\lambda_{T}}{a}$ for fermions (black). The dashed (red) line shows the same result obtained with the expression of the weak coupling expansion.}
\label{wfb3}
\end{figure}

Figure \ref{wbb2} shows the second virial coefficient for bosons against $\alpha$ in the weak coupling approximation and the numerical result obtained in the latest section, Figure \ref{wfb2} does the same for the fermionic situation and Figures \ref{wbb3} and \ref{wfb3} do the same for the third coefficient of bosons and fermions respectively. It is easy to see that the curves corresponding to equations (\ref{wb2s}) and (\ref{wb3s}) and the ones obtained  numerically  integrating the expressions  (\ref{b2s}) and (\ref{b3s}),  shown  in figures \ref{wbb2},\ref{wfb2},\ref{wbb3} and \ref{wfb3},  are almost indistinguishable for $|\alpha|>4$.  

\section{virial coefficients for trapped gases}

\qquad

In order to study  the influence of a  harmonic trap for quantum gases it is possible to use the local density approximation (LDA). The LDA may be used if one ignores the variation of thermodynamic quantities due to density gradients \cite{DalfovoI,DalfovoII}.   In our formalism this  means that one can replace the chemical potential by $\mu \rightarrow \mu - V(\mathbf{r})$, giving a free energy $F(\mathbf{r})$ that depends on  $\mathbf{r}$.   The final free energy will be given by: $F=\int F(\mathbf{r}) d^3 \mathbf{r}$. 

We know that the virial cofficients are related to the free energy by equation 
\ref{virial}.  
Therefore, in  the LDA approximation, expression (\ref{virial}) becomes:
\begin{equation} \label{F_LDA}
F=- \frac {1} {\beta \lambda^3} \sum_{n=1}^{\infty} 
\left(b_n \int e^{-\beta n V(\mathbf{r})} d^3 \mathbf{r} \right) z^n.
\end{equation}
Comparing (\ref{F_LDA}) to (\ref{virial}) one sees 
that the presence of the trap changes the virial coefficients in the following way under the LDA approximation:
\begin{equation}
b_n \rightarrow b_n \int e^{-\beta n V(\mathbf{r})} d^3 \mathbf{r}.
\end{equation}

Considering an anisotropic harmonic trap $$V(\mathbf{r})=\sum_{i=1}^{3} \left[ \frac {w_i x_{i}^{2}} {2} \right] $$ we obtain 
\begin{equation} \label{virial_aniso}
b_n \rightarrow  b_n \left( \frac {2 \pi} {\beta n} \right)^{\frac {3} {2}}. \left[ \prod_{i=1}^{3} w_i \right]^{-\frac {1} {2}}.
\end{equation}
In particular, if the trap is isotropic,  $w_1=w_2=w_3=w$,  we arrive at the following result:
\begin{equation}  \label{virial_iso}
b_n \rightarrow b_n \left( \frac {2 \pi} {\beta w n} \right)^{\frac {3} {2}}.
\end{equation}
One sees that for $w$ given by the fundamental Matsubara frequency $\frac{2 \pi} {\beta}, $ the first virial coefficient $b_{1}$  does not change in the presence of the harmonic isotropic trap.
In the following figures we plot the ratios $b_{n}'/b_{n}$ for $n=2, 3, 4$, where $'$  means the presence of the harmonic isotropic trap, as a function of  $w$ for $T=1$,  and  also as
a function of  $T$ for $w=1$.

\begin{figure}[tbp]
\centering    
\includegraphics[width=0.5\textwidth]{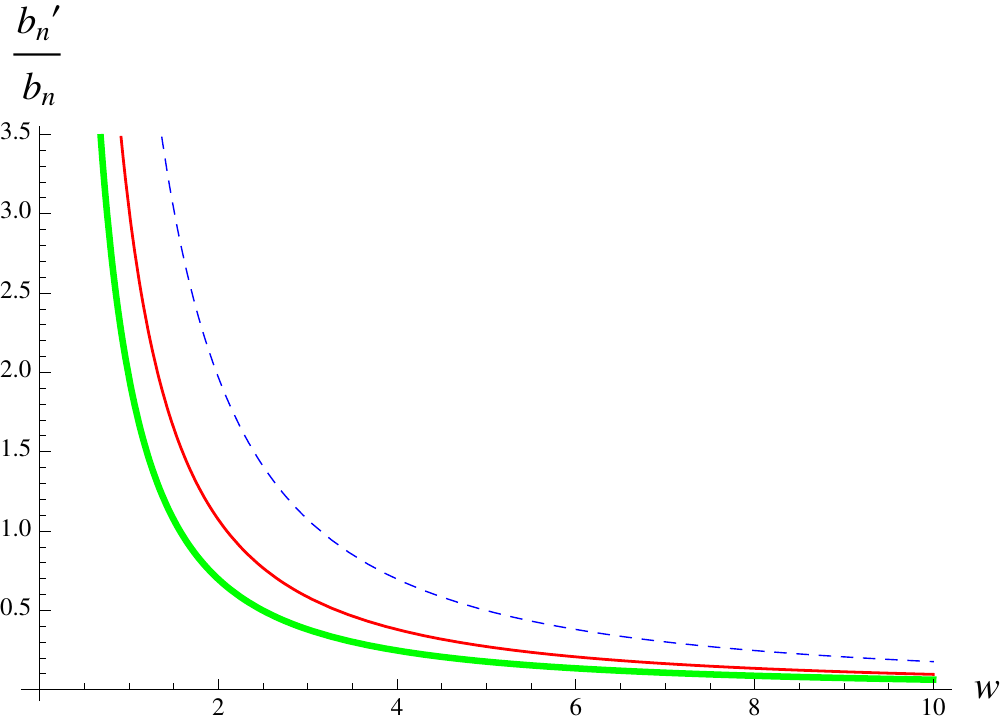}
\caption{(Color Online): Ratio between the second, third and fourth virial coefficients in the presence and in the absence of an harmonic isotropic trap against the frequency of the trap for $T=1$ in the three-dimensional case ($d=3$),  $\frac{b_{n}'} {b_{n}} \times w$.  $n=2$ is the dashed (blue) line, $n=3$ is the thin (red) line and $n=4$ is the thick (green) line. }
\label{LDAxW}
\end{figure}

\begin{figure}[tbp]
\centering    
\includegraphics[width=0.5\textwidth]{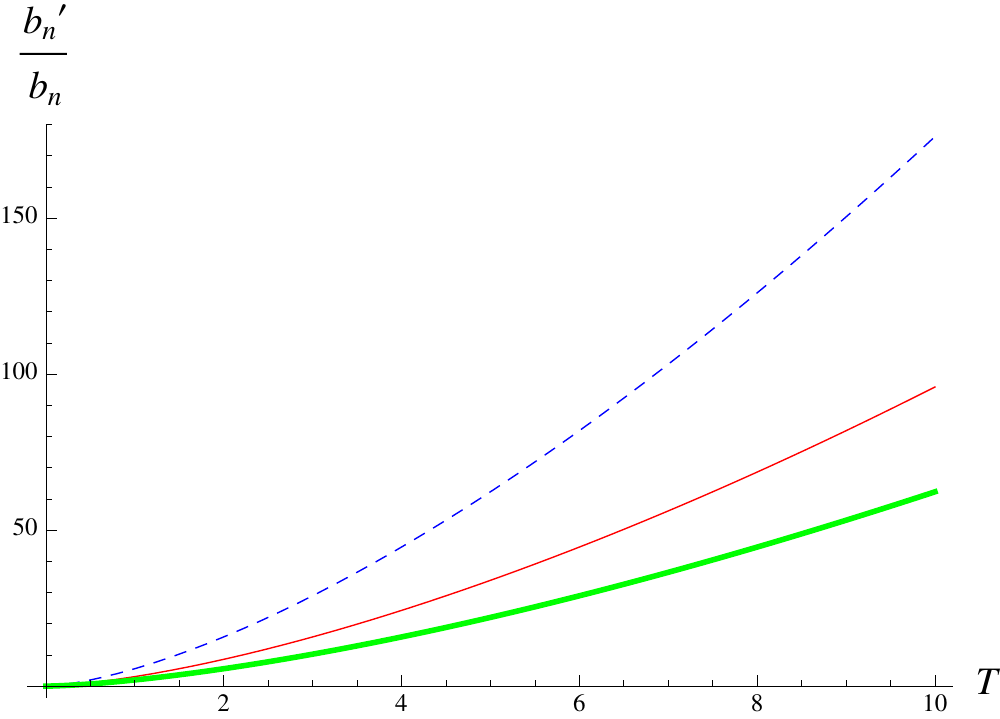}
\caption{(Color Online): Ratio between the second, third and fourth virial coefficients in the presence and in the absence of an harmonic isotropic trap against  the temperature for $w=1$ in the three-dimensional case ($d=3$), $\frac{b_{n}'} {b_{n}} \times T$.  $n=2$ is the dashed (blue) line, $n=3$ is the thin (red) line and $n=4$ is the thick (green) line.  }
\label{LDAxT}
\end{figure}

Expressions (\ref{virial_aniso}) and (\ref{virial_iso}) show that the virial coefficients decrease monotonically with the frequencies of the trap and increase monotonically with the temperature as a power law as Figures (\ref{LDAxW}) and (\ref{LDAxT}) show for the isotropic case.

\section{Conclusions}

\qquad

The first virial coefficients of a bosonic and a fermionic gas were obtained as functions of the ratio of the thermal wave length to the scattering length $\alpha=\frac{\lambda_{T}}{a}$ in the foam diagram approximation. The results obtained in \cite{Out_unitarity} are recovered when $\alpha \rightarrow 0^{\pm}$ as expected and one also recovers  the free case when $\alpha \rightarrow \pm \infty$.

The second virial coefficients are exact and the unitary limit values for the fermionic case agree with the  results of \cite{HoMu} as is explained in \cite{Out_unitarity}. The third coefficient is not  exact since the foam diagram approximation neglects  3-body interactions,   however  it becomes very close to the correct value when the absolute value of the ratio $\alpha$ is large.
 
A weak coupling expansion was performed and analytical expressions for  the virial coefficients were  obtained for large values of $|\alpha|$.   The weak coupling expansion 
is in close agreement with the results we obtained for any $\alpha$  when $|\alpha|>4$.

The influence of an anisotropic harmonic trap was also  considered under the Local Density Approximation and analytical expressions were obtained,  and also specialized to 
an isotropic trap.

\begin{acknowledgments} 
The authors acknowledge financial support from CNPq, CAPES, and FAPERJ. This work is partly funded by a Science Without Borders-CNPq grant. 
\end{acknowledgments} 


\begin{thebibliography}{99}


\bibitem{Cold_atoms1} Y. Shin, C. Schunck, A. Schirotzek and W. Ketterle,                  
{\it Phase diagram of a two-component Fermi gas with resonant interactions},
Nature 451 (2008) 689.

\bibitem{Cold_atoms2} I. Bloch, J. Dalibard and W. Zwerger, 
{\it Many-body physics with ultracold gases},
Rev. Mod. Phys. 80 (2008) 885.

\bibitem{Cold_atoms3} C. Chin, R. Grimm, P. Julienne and E. Tiesinga, 
{\it Feshbach Resonances in Ultracold Gases}, 
Rev. Mod. Phys. 82 (2010) 1225. 

\bibitem{Cold_atoms4} S. Nascimb\`ene, N. Navon, K. J. Jiang, F. Chevy and C. Salomon, 
{\it Exploring the thermodynamics of a universal Fermi gas}, 
NATURE 463 (2010) 1057.

\bibitem{Cold_atoms5} K. Van Houcke, F. Werner ,E. Kozik,  N. ProkofÕev, B. Svistunov, M.J.H. Ku, A.T.Sommer, L.W. Cheuk,
 A. Schirotzek, 	 and M.W. Zwierlein, 
{\it Feynman diagrams versus Fermi-gas Feynman emulator},
Nature Physics (2012) doi:10.1038/nphys2273 


\bibitem{Cold_atoms6} M. Horikoshi, S. Nakajima, M. Ueda and T. Mukaiyama, 
{\it Measurement of universal thermodynamic functions for a unitary Fermi gas}, 
Science 327 (2010) 442.

\bibitem{MC1} A. Minguzzi, S. Succi, F. Toschi, M.P. Tosi and P. Vignolo,
{\it Numerical methods for atomic quantum gases with applications to Bose-Einstein condensates and to ultracold fermions},
Phys. Reports 395 (2004) 223.

\bibitem{MC2}   E.  Burovski,  N.  Prokof'ev,  B. Svistunov and M.  Troyer, 
{\it Critical temperature and thermodynamics of attractive fermions at unitarity},  Phys. Rev. Lett {\bf 96} (2006)  160402.

\bibitem{MC3} A. Bulgac, J. E. Drut and P. Magierski,
{\it Quantum Monte Carlo Simulations of the BCS-BEC Crossover at Finite Temperature}, Phys.Rev. A78 (2008) 023625. 

\bibitem{MC4} P. Magierski, G. Wlazlowski, A. Bulgac and J.E. Drut,
{\it Finite-Temperature Pairing Gap of a Unitary Fermi Gas by Quantum Monte Carlo Calculations}
Phys. Rev. Lett. 103 (2009) 210403.

\bibitem{Review1}   Y. Castin and F.  Werner,  
{\it The Unitary Gas and its Symmetry Properties}, 
Springer Lecture Notes in Physics,  
``BEC-BCS  Crossover and the Unitary Fermi Gas'',  
W. Zwerger, ed.   [arXiv:1103.2851]

\bibitem{Review2}   H. Hu,  X.-J. Liu and P. D. Drummond, 
{\it Universal thermodynamics of a strongly interacting Fermi gas:  theory verses experiment}, 
New J. of Phys. {\bf 12}, 063038 (2010).  

\bibitem{Harmonic} J.R. Armstrong, N.T. Zinner, D.V. Fedorov, A.S. Jensen, {\it Virial expansion coefficients in the harmonic approximation},
Phys. Rev. E 86, 021115 (2012).

\bibitem{Dashen}   R. Dashen, S-K. Ma, Herbert J. Bernstein,
{\it S-matrix formulation of Statistical Mechanics},
Phys. Rev. 187, 1 (1969).

\bibitem{YangYang}    C. N. Yang and C. P.  Yang,  Jour. Math.  Phys. {\bf 10}  (1969)  1115.  

\bibitem{PyeTon}   P. T. How and A.  LeClair, 
{\it Critical point of the two-dimensional Bose gas: an S-matrix approach},
Nucl. Phys. B824 (2010) 415 [arXiv:0906.0333]

\bibitem{PyeTonUnitary1}   P.-T. How and A. LeClair,  
{\it S-matrix approach to quantum gases in the unitary limit I:
 the two-dimensional 
case},  
J.Stat.Mech. (2010)  P03025.


\bibitem{PyeTonUnitary2}   P.-T. How and A. LeClair, 
{\it S-matrix approach to quantum gases in the unitary limit II:
 the three-dimensional 
case},  
J. Stat. Mech. (2010) P07001. 

\bibitem{Out_unitarity}  A. LeClair, E. Marcelino, A. Nicolai and I. Roditi,
 {\it Quantum Bose and Fermi gases with large negative scattering length},
Phys. Rev. A 86, 023603 (2012). [arXiv: 1205.0234]

\bibitem{Leclair_viscosity}  A. LeClair,
 {\it On the viscosity-to-entropy density ratio for unitary Bose and Fermi gases},
New J. of Phys. 13 (2011) 055015.

\bibitem{Exp_viscosity}   C.  Cao,  E.  Elliot,  J.  Joseph,  H.  Wu,   J.  Petricka,  T. Sch\"afer and J. E. Thomas,  
{\it  Universal Quantum Viscosity in a Unitary Fermi Gas},  
Science {\bf 331} (2011)  58. 


\bibitem{HoMu}  T. L. Ho and E. J. Mueller,  
{\it High Temperature Expansion Applied to Fermions near Feshbach Resonance},
Phys. Rev. Lett.  {\bf 92}  (2004)  160404.

\bibitem{b3Theory}   A.J. Liu,  H.  Hu and P. D. Drummond, 
{\it  Virial expansion for a strongly correlated Fermi gas},   
  Phys.  Rev. Lett. {\bf 102} (2009)  160401. 
  
\bibitem{Leyronas}   X.  Leyronas,  
{\it Virial expansion with Feynman diagrams}, 
 Phys. Rev. {\bf A84}  (2011)  053633.   
 
 \bibitem{Kaplan}   D. B.  Kaplan and S.  Sun, 
 {\it A new field theoretic method for the virial expansion}
 Phys.  Rev.  Lett. {\bf 107}  (2011)  030601. 
 
 \bibitem{Castin}   Y. Castin and F\'elix Werner, 
  {\it Third virial coefficient for the unitary Bose gas,}
  Canadian. Jour. Phys.  {\bf 91} (2013)  382. 
  
 \bibitem{Up1} K. Dieckmann, C.A. Stan, S. Gupta, Z. Hadzibabic, C.H. Schunck, and W. Ketterle, 
{\it Decay of an ultracold fermionic lithium gas near a Feshbach resonance},
Phys. Rev. Lett. 89 (2002) 203201. 

\bibitem{Up2} S. Jochim, M. Bartenstein, A. Altmeyer, G. Hendl, C. Chin, J.H. Denschlag and R. Grimm,
{\it Pure gas of optically trapped molecules created from fermionic atoms} 
Phys. Rev. Lett. 91 (2003) 240402. 

\bibitem{Up3} J. P. Gaebler, J. T. Stewart, T. E. Drake, D. S. Jin, A. Perali, P. Pieri and G. C. Strinati,  
{\it Observation of pseudogap behaviour in a strongly interacting Fermi gas},
Nature Physics 6 (2010) 569.

\bibitem{Up4} S. Tsuchiya, R. Watanabe, and Y. Ohashi, 
{\it Single-particle properties and pseudogap effects in the BCS-BEC crossover regime of an ultracold Fermi gas above $T_c$},
Phys. Rev. A 80, 033613 (2009).

\bibitem{Up5} V. B. Shenoy and T.-L. Ho, 
{\it Nature and Properties of a Repulsive Fermi Gas in the Upper Branch of the Energy Spectrum},
Phys. Rev. Lett. {\bf 107}   (2011) 210401.

\bibitem{Up6} Density and spin response of a strongly-interacting Fermi gas in the attractive and quasi-repulsive regime,  F. Palestini, P. Pieri, G. C. Strinati, Phys. Rev. Lett. {\bf 108}  (2012) 080401.

\bibitem{Up7} W. Li, T-L. Ho, {\it Bose Gases Near Unitarity},
arXiv:1201.1958v3 [cond-mat.quant-gas].
 
\bibitem{Ketterle}   W.  Ketterle,  private communication.  

\bibitem{DalfovoI} F. Dalfovo and S. Giorgini, {\it Theory of Bose-Einstein condensation in trapped gases},
Rev. Mod. Phys. 71 (1999) 463.

\bibitem{DalfovoII} F. Dalfovo and S. Stringari, {\it Bosons in anisotropic traps: Ground state and vortices},
Phys. Rev. A 53, 2477 (1996).

\bibitem{LL} L. D. Landau and E. M. Lifshitz,
{\it Quantum Mechanics: Non-relativistic Theory}, Pergamon, Oxford, 1977.



 



\end{thebibliography}
\end{document}